%% file: main.tex
\documentclass[sigconf,screen=true,bookmarks=false]{acmart}

\AtBeginDocument{%
  }

\setcopyright{acmlicensed}
\copyrightyear{2018}
\acmYear{2018}
\acmDOI{XXXXXXX.XXXXXXX}

\acmConference[ASP-DAC 2025]{Make sure to enter the correct
  conference title from your rights confirmation emai}{Jan. 20--23,
  2025}{Tokyo, Japan}
\acmISBN{978-1-4503-XXXX-X/18/06}

\usepackage[]{algpseudocode}
\usepackage{xspace}
\usepackage{balance}
\algdef{SE}[DOWHILE]{Do}{doWhile}{\algorithmicdo}[1]{\algorithmicwhile\ #1}%

\input{packages}
\input{macro}
\graphicspath{{./figs/}}

\begin{document}
\settopmatter{printacmref=false} %
\renewcommand\footnotetextcopyrightpermission[1]{} %
\pagestyle{plain} %

\title{
APR: Automated Photonic Integrated Circuit Detailed Routing with Curvy Waveguide and Adaptive Crossing Insertion
}

\title{
Automated Curvy Waveguide Routing\\ for Large-Scale Photonic Integrated Circuits
}

\author{
Hongjian Zhou$^1$,
Keren Zhu$^2$,
Jiaqi Gu$^{1\dagger}$\\
$^1$Arizona State University, $^2$Fudan University \\ {\small $\dagger$jiaqigu@asu.edu}
}

\input{doc/1_abstract}
\maketitle
\input{doc/2_intro}

\input{doc/3_prelim}

\input{doc/4_algo}

\input{doc/5_result}

\input{doc/6_conclu}


\input{main.bbl}
\end{document}

%% file: packages.tex
\usepackage[utf8]{inputenc} %
\usepackage[T1]{fontenc}    %
\usepackage{pifont}
\usepackage{xcolor}         %
\usepackage{booktabs}
\usepackage{hyperref}
\hypersetup{colorlinks,
    linkcolor=red,citecolor=citecolor,urlcolor=blue}
\usepackage{algorithm}
\usepackage{graphicx}
\usepackage{threeparttable}
\usepackage{multirow}
\usepackage{amsmath}
\usepackage{bm}
\usepackage{bbm}
\usepackage{amsfonts}
\usepackage{amsthm}
\usepackage[]{algpseudocode}
\usepackage{enumitem} %
\usepackage[subrefformat=parens,labelformat=parens]{subfig}
\usepackage{colortbl}

\usepackage{wrapfig}
\usepackage[algo2e, ruled, vlined, linesnumbered, noend]{algorithm2e}

\newenvironment{problem}[1][Problem]
  {\begin{trivlist} \item[\hskip \labelsep {\bfseries #1.}]}
  {\end{trivlist}}

\setcitestyle{sort&compress,numbers}

%% file: macro.tex
\definecolor{citecolor}{RGB}{34,139,34}
\definecolor{mydarkblue}{rgb}{0,0.08,1}
\definecolor{mydarkgreen}{rgb}{0.02,0.6,0.02}
\definecolor{mydarkred}{rgb}{0.8,0.02,0.02}
\definecolor{mydarkorange}{rgb}{0.40,0.2,0.02}
\definecolor{mypurple}{RGB}{111,0,255}
\definecolor{myred}{rgb}{1.0,0.0,0.0}
\definecolor{mygold}{rgb}{0.75,0.6,0.12}
\definecolor{myblue}{rgb}{0,0.2,0.8}
\definecolor{mydarkgray}{rgb}{0.,0.2,0.2}

\definecolor{lightred}{RGB}{255,235,235}
\definecolor{lightgreen}{RGB}{235,255,235}
\definecolor{lightblue}{RGB}{235,235,255}
\definecolor{lightcyan}{RGB}{235,255,255}
\definecolor{lightmagenta}{RGB}{255,235,255}
\definecolor{lightyellow}{RGB}{255,255,235}

\definecolor{qxkcolor}{RGB}{215,235,255}
\definecolor{softmaxcolor}{RGB}{230,235,255}
\definecolor{probxvcolor}{RGB}{255,255,235}

\definecolor{topkcolor}{RGB}{255,235,235}
\definecolor{zecolor}{RGB}{255,255,235}
\definecolor{dynacolor}{RGB}{235,255,255}

\definecolor{reviewcolor}{RGB}{0,0,200}

\newcommand{\ceil}[1]{\lceil #1 \rceil}

\theoremstyle{plain}

\theoremstyle{definition}

\algdef{SE}[DOWHILE]{Do}{doWhile}{\algorithmicdo}[1]{\algorithmicwhile\ #1}%

\newcommand{\squishlist}{
 \begin{list}{$\bullet$}
  { \setlength{\itemsep}{0pt}
     \setlength{\parsep}{3pt}
     \setlength{\topsep}{3pt}
     \setlength{\partopsep}{0pt}
     \setlength{\leftmargin}{1.5em}
     \setlength{\labelwidth}{1em}
     \setlength{\labelsep}{0.5em} } }

\newcommand{\squishend}{
  \end{list}  }

\newcommand{\name}{\texttt{APR}\xspace}

%% file: doc/1_abstract.tex
\begin{abstract}
\label{abstract}
As photonic integrated circuit (PIC) designs advance and grow in complexity, largely driven by innovations in photonic computing and interconnects, traditional manual physical design processes have become increasingly cumbersome.
Available PIC layout automation tools are mostly schematic-driven, which has not alleviated the burden of manual waveguide planning and layout drawing for engineers.
Previous research in automated PIC routing largely relies on off-the-shelf algorithms designed for electrical circuits, which only support high-level route planning to minimize waveguide crossings.
It is not customized to handle unique photonics-specific routing constraints and metrics, such as curvy waveguides, bending, port alignment, and insertion loss.
These approaches struggle with large-scale PICs and cannot produce real layout geometries without design-rule violations (DRVs).
This highlights the pressing need for electronic-photonic design automation (EPDA) tools that can streamline the physical design of modern PICs.
In this paper, \emph{for the first time}, we propose an open-source automated PIC detailed routing tool, dubbed \name, to generate DRV-free PIC layout for large-scale real-world PICs.
\name features a grid-based curvy-aware A$^\ast$ engine with adaptive crossing insertion, congestion-aware net ordering and objective, and crossing-waveguide optimization scheme, all tailored to the unique property of PIC.
On large-scale real-world photonic computing cores and interconnects, \name generates a DRV-free layout with 14\% lower insertion loss and 6.25$\times$ speedup than prior methods, paving the way for future advancements in the EPDA toolchain.
Our codes are open-sourced at \href{https://github.com/ScopeX-ASU/APR}{link}.

\end{abstract}

%% file: doc/2_intro.tex
\section{Introduction}
\label{sec:Introduction}

In recent years, as silicon photonics advances, photonic integrated circuits (PICs) have received significant attention among researchers due to the characteristics of high-speed and low-power dissipation.
There are various designs and demonstrations on photonic tensor cores (PTCs) for optical neural networks (ONNs), and photonic network-on-chips (NoCs) for high-bandwidth chip communications.
Due to those main driving research areas, PICs exhibit an exponential increase in complexity. 
As shown in Fig.~\ref{fig:layout}, the number of photonic components on a single chip is rapidly approaching the order of 1000 components per chip and is expected to double every 2.5 years~\cite{thylen2006moore}.
There is an increasing demand for electronic-photonic design automation (EPDA) toolkits to automate layouts, improving both productivity and solution quality.

\input{figtex/fig_pdflow}

Traditionally, PIC physical design is schematic-driven~\cite{korthorst2023photonic}. 
In this approach, components are placed and interconnected according to the circuit topology and signal paths in the schematic, aiming to minimize crossings, detours, and bending, which helps reduce insertion loss and improves signal integrity.
In certain cases, routing can be \textbf{manually managed}, particularly when the circuits are \emph{highly structured with a well-designed no-crossing topology}, such as in crossbar arrays~\cite{NP_Nature2021_Feldmann} or triangular/rectangular meshes~\cite{NP_NATURE2017_Shen}, binary tree structure~\cite{hashemi2022review}
When these designs are \emph{optimally placed with large spacing and perfectly aligned ports}, device abutment or simple straight waveguides can automatically connect ports, similar to the standard cell-based layout used in SRAM arrays.

However, significant routing challenges arise, requiring PIC routing automation when:
\ding{202}~the circuit \textbf{scale exceeds manual capabilities}, such as with hundreds or thousands of instances/nets;
\ding{203}~the circuits have \textbf{complicated topology or are not perfectly placed}, leading to issues like port offsets, numerous crossings, and routing congestion;
\ding{204}~the design needs to \textbf{adapt to different fabrication processes or device designs}, each with varying component sizes and properties, necessitating adjustments in waveguide routing;
\ding{205}~frequent updates or design iterations make manual modifications inefficient, particularly when schematic designers \textbf{lack full visibility into waveguide routing and where crossings need to be inserted}. 
This often leads to repeated back-and-forth between schematic and layout design, especially during layout design space exploration with iterative placement and routing.

Most of the existing work focuses on \underline{PIC global route planning}.
Optical routing algorithms~\cite{minz2007optical, ding2009oil, ding2009router} are proposed for on-chip 3D system-on-package designs, primarily aiming at optimizing the signal loss and total power.
PROTON~\cite{proton} and PLATON~\cite{von2016platon} are automatic place-and-route tools where a modified Lee’s algorithm is used for optical waveguide routing. 
In ~\cite{chuang2018planaronoc}, the insertion loss is further reduced by optimizing device flipping and rotation to minimize crossings.
Those existing global routing approaches primarily \textbf{focus on planning/finding paths} that minimize path loss, but often \textbf{overlook the physical implementation} of these paths. 
This can result in issues like \emph{path congestion}, \emph{failure to insert crossings or bends}, ultimately leading to an \emph{invalid routing solution}.

There is also work focused on completing the \underline{detailed routing} stage. 
Prior work~\cite{condrat2012methodology} solved global routing using mixed integer programming, followed by a Manhattan grid-based detailed routing using a left-edge algorithm, with crossings treated as design constraints. 
However, this grid-based approach limits waveguide bends to 90$^\circ$.
A subsequent approach~\cite{condrat2013channel} introduced non-Manhattan channel routing to handle optical waveguide curves. Yet, since current PICs typically use only a single optical waveguide layer, \emph{crossings are inevitable}, leaving little room for optimization during detailed channel routing.
To address these limitations, it is essential to have an automated PIC router that is \textbf{fully aware of the physical instantiation and design rule} of waveguides and components with smart crossing insertion.

In this work, we propose an automated PIC detailed routing tool featuring non-Manhattan curvy waveguide handling and adaptive crossing insertion. 
Our framework addresses key limitations of existing methods by not only optimizing path insertion loss but also considering waveguide geometry and layout constraints during routing. 
By \textbf{adaptively inserting crossings}, rather than manually pre-inserting them in the schematic, and \textbf{considering the actual geometry} of waveguides, crossings, and bends, \name can generate \textbf{complete and design-rule violation (DRV)-free layout in minutes}, minimizing the need for extensive post-routing adjustments or iterative schematic/layout modification.

Highlights of this work are summarized as follows.
\begin{itemize}
  \item We devise a fully automated PIC detailed routing tool \name that delivers DRC-free, low insertion loss routed PIC layout for large-scale circuits in minutes, supporting curvy waveguide geometry and automatic crossing insertion.  
  \item \textbf{Curvy-Aware Non-Manhattan A$^\ast$ Router}: A customized curvy-aware A$^\ast$ search with adaptive neighbors to support different types of curvy structures. 
  \item \textbf{Accessibility-Enhanced Port Assignment}: We introduce synergistic strategies to improve PIC routability by enabling orientation-aware port access and reserving space in port-congested regions.
  \item \textbf{Congestion-Penalized RR with Grouped Net Order}: We propose group-based net order, group congestion penalty, and local rip-up \& reroute (RR) strategies to minimize waveguide crossings for better routability.
  \item On large-scale PTC and oNoC benchmarks, our proposed \name generates DRV-free layouts with 14\% lower insertion loss (dB) and 6.25$\times$ faster runtime than prior methods.
\end{itemize}

%% file: figtex/fig_pdflow.tex
\begin{figure}
    \centering
    \includegraphics[width=0.95\columnwidth]{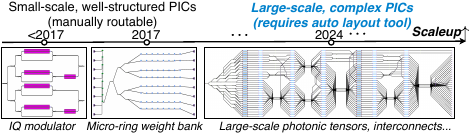}
    \vspace{-5pt}
    \caption{Modern PIC scale and complexity require EPDA.}
    \label{fig:layout}
     \vspace{-5pt}
\end{figure}

%% file: doc/3_prelim.tex
\section{Preliminaries}
\label{sec:Preliminaries}

In this session, we will first give a brief background on related VLSI routing methods, and the differences between PIC routing and VLSI routing by discussing PIC design rules. Following that, we will outline the traditional manual PIC routing flow. 
Finally, we will present the evaluation metrics for PIC routing and define the specific challenges associated with it.
The notations used in this paper is summarized in Table~\ref{tab:notation}.
\input{tables/tab_notation}

\subsection{VLSI Detailed Routing}

Detailed routing faces challenges such as complex design rules, pin access, and limited routing resources~\cite{Routing_ISPD22_Posser}. 
Common VLSI routing methods, for example~\cite{Routing_TCAD22_Kahng, Routing_DAC12_Gester, Routing_ICCAD19_Li, Routing_ICCAD18_Sun},utilize path-finding algorithms such as A$^\ast$ search or maze routing, supported by a DRC engine. 
A classic and still popular paradigm for resolving competition over routing resources between different nets is \textit{negotiation}-based routing~\cite{Routing_FPGA95_McMurchie}. In these approaches, a rip-up and reroute scheme is utilized to clear routing failures.

A key distinction between VLSI and PIC routing is the routing direction.
VLSI routing is usually Manhattan or even unidirectional, while PIC routing necessitates curvy waveguides. 
Diverse routing directions, including octagonal routing, are employed in analog~\cite{Routing_Access23_Martins, Routing_TCAD14_Ou}, PCB~\cite{Routing_ICCAD10_Yan, Routing_DAC23_Liu, Routing_DAC21_Lin}, and package routing~\cite{Routing_DAC23_Chung, Routing_ICCAD23_Lee}. 
Nevertheless, research on curvy path routing remains limited.

\subsection{Photonic Design Rules}
In the PIC routing problem, we typically operate with a single silicon waveguide routing layer, where photonic devices are considered as obstacles. 
The waveguides form port-to-port optical paths, resulting in all nets being 2-pin nets.
Here, we provide a brief overview of PIC routing design rules and highlight the unique considerations specific to photonic circuits.
\input{figtex/fig_eocomparison}

\subsubsection{Waveguide Spacing}
Waveguides need proper spacing with each other and photonic device structures to avoid crosstalk from unwanted coupling.
Due to the diverse types/sizes of waveguides, as shown in Fig.~\ref{fig:EOComparison}, the minimum spacing rule between two nets depends on many factors, e.g., wavelength, polarization mode, refractive index contrast, substrate type, waveguide cross-sections.
For example, 
for high-index contrast systems (such as silicon-on-insulator), small spacings (e.g., 1-3 $\mu$m) are sufficient.

\subsubsection{Bend Radius}
In photonic circuits, the bend radius is a key parameter that has a huge difference from the 90$^\circ$ metal wire bend in VLSI.
Sharp bends in photonic waveguides can cause significant mode mismatch and radiation losses. 
To mitigate these losses, the waveguide bend structure typically forms a smooth curve, such as a circular or Euler bend for 90$^\circ$ turns and a sine bend for routing offset, with sufficient curvature to ensure proper light confinement and minimize loss, as shown in Fig.~\ref{fig:EOComparison}.

The bend radius in photonic circuits can vary widely, typically ranging from a few microns to millimeters, depending on factors like material properties, bending structure, and refractive index contrast. 
Silicon waveguides with high refractive index contrast can support small bend radii, typically around 5-10 $\mu m$.
Silicon nitride waveguides, with lower refractive index contrast, require larger bend radii, generally 20-100 $\mu m$, depending on waveguide geometry and application.
While a larger bend radius minimizes insertion loss, it also consumes more chip area and routing resources, which can potentially cause routability issues.

\subsubsection{Waveguide Crossing}
Unlike VLSI routing that forbids wire crossings and uses vias for layer transitions, photonic circuits enable waveguide crossings (CRs) on the same layer.
CRs are often essential especially for dense circuits.
However, each CR introduces insertion loss, typically ranging from 0.1 dB to 1 dB, and occupies a footprint of about $\sim$ 5$\times$5 $\mu m^2$.
Moreover, the angle at which waveguides intersect is crucial in minimizing crosstalk.
CRs require \emph{perpendicular waveguide intersections to minimize crosstalk}, posing \emph{challenges for routing dense PICs}, especially when parallel waveguides need sufficient space to adjust their relative orientation through curvy bending, as shown in Fig.~\ref{fig:EOComparison}.

\subsubsection{Port Connection and Alignment}
PIC connects waveguides via precise port abutment, which requires exact face-to-face alignment (180$^\circ$ orientation).
Fig.~\ref{fig:EOComparison} shows an example.
Misalignment or offset between waveguides can lead to signal path failure, making \emph{precise alignment} a critical requirement during PIC routing.

\subsubsection{Signal Integrity}
One of the most important metrics for PIC is insertion loss, which impacts the laser power budget and signal integrity (signal-to-noise ratio, crosstalk).
The major evaluation metric for PIC routing is the maximum insertion loss on the critical path.
Long waveguides and CRs introduces disturbance in signal integrity and are preferred to be avoided.

\subsection{Schematic-Driven PIC Layout}

Traditional PIC physical design workflows, including manual design and current available EPDA tools, are schematic-driven~\cite{chrostowski2016schematic}.
In this approach, all structures, including crossings and even each segment of a waveguide, are treated as separate instances in the netlist.
Designers need to plan the routing ahead during schematic stage and manually insert crossings as instances to the netlist.
Then, the nets in the schematic represent only port connectivity, eliminating the need for physical instantiation of nets, as all ports of waveguides are connected through abutment. 

One significant drawback of the schematic-driven layout approach is that waveguide routing and crossings must be predetermined by design experts at the schematic stage, relying on empirical predictions of physical design solutions. 
Once established, these elements cannot be easily added or removed during routing, resulting in a \textbf{rigid, manually-defined routing topology}. 
This rigidity often causes back-and-forth modifications between the physical design and schematic stages, which can be inefficient.
Moreover, it is \textbf{not scalable} to manually handle the routing of large-scale PICs.
To address this issue, a \textbf{new formulation of instances and nets is needed to decouple the schematic and physical design stages} while incorporating automated crossing insertion. 
This would allow for \textbf{greater flexibility and efficiency for scalable} PIC auto-routing.

\subsection{PIC Routing Quality Metrics}
In addition to regular routing metrics,  such as wirelength, design rule violation, and runtime, one of the most important photonic-specific metrics is \textbf{critical path insertion loss (IL)} that impacts link power budget and signal-to-ratio ratio.
IL is calculated based on the optical path which refers to the light propagation path through all cascaded components from the laser source to the photodetector. 
Assume a path $p_i$ consists of instances and nets $(m_0\rightarrow n_0\rightarrow m_1\rightarrow n_1 \rightarrow \cdots)$. 
Some nets and instances are shared across different paths.
Note that for multi-port photonic devices, we assume the same IL from any input port to any output port given lack of accurate IL information from available free PDKs.
Port-specific ILs can be easily considered in the same formulation.
The insertion loss $IL(p_i)$ is defined as the sum of ILs of all devices $IL(m_j)$ and waveguide routes $IL(n_j)$ along the path in the decibel unit (dB) as a convention. 
For net IL, we will consider the crossing $IL_{cr}$, bending $IL_{bn}$, and propagation $IL_{wg}$ losses in the instantiated waveguide routes. 
Therefore, we have:
\begin{equation} 
\label{eq:iloss}
\small
\begin{aligned}
    &IL(p_i) = \!\!\!\sum_{m_j\in p_i}IL(m_j)+\sum_{n_j\in p_i} IL(n_j)\\
    \sum_{n_j\in p_i}&IL(n_j)=IL_{wg}(p_i) + IL_{cr}(p_i) + IL_{bn}(p_i)\\
    IL_{wg}(p_i) = \alpha_w&  WL_{p_i}, \ 
    IL_{cr}(p_i) = \alpha_c  \# CR_{p_i}, \ 
    IL_{bn}(p_i) = \alpha_b  \angle BN_{p_i},
\end{aligned}    
\end{equation}
where $WL_{p_i}$, $\# CR_{p_i}$, and $\angle BN_{p_i}$ are the total straight waveguide length, the number of crossings and total degree of bending along the path $p_i$, and coefficients $\alpha_w, \alpha_c, \alpha_b$ are the insertion loss per unit length/CR/angle for the specific photonic component structures.
To achieve the desired optical functionality and signal-to-noise ratio for switching, modulation, or multiplexing, the insertion loss should be minimized. 

The maximum insertion loss $IL_{max}$ among all paths $P$ determines how much extra power is required from the laser to ensure that enough light reaches the output photodetectors or subsequent stages in the circuit. 
Thus, $IL_{max}$ is the main qualifier of PIC routing, and the objective function is given as:
\begin{equation} \label{eq:ilmax}
IL_{max} = max_{p_i \in P}~~IL(p_i)
\end{equation}

\subsection{Problem Formulation} \label{subsec:problem_formulation}
We formally define the PIC detailed routing problem as follows.
\begin{problem}[PIC Detailed Routing]
\textit{Given a set of nets $N = \{n_i | 1 \leq i \leq |N| \}$, a set of placed devices $M = \{m_i | 1 \leq i \leq |M|\}$, generate a routing solution for each net $n_i \in N$ such that $n_i$ is connected without design rule violations and minimize the $IL_{max}$.}
\end{problem}

%% file: tables/tab_notation.tex
\begin{table}
\centering
\caption{Notations used in this paper.
}
\vspace{-10pt}
\resizebox{8cm}{!}{
\begin{tabular}{|c|p{6.8cm}|}
\hline
Symbol & Description \\ \hline
$N$                & The set of nets specified in the circuit netlist. \\
$n_i$              & The $i^\mathrm{th}$ net in $N$, $1 \leq i \leq |N|$. \\
$P$                & The set of all paths. \\
$p_i$              & The $i^\mathrm{th}$ path in $P$, $1 \leq i \leq |P|$. \\
$IL(p_i)$          & The insertion loss of the $p_i$. \\
$IL_{max}$         & The maximum insertion loss over all paths. \\
$IL_{wg}(p_i)$     & The propagation loss of the path. \\
$IL_{cr}(p_i)$     & The crossing loss of the path. \\
$IL_{bn}(p_i)$     & The bending loss of the path. \\
$\alpha_w$, $\alpha_c$, $\alpha_b$        & Coefficient of $IL_{wg}(p_i)$, $IL_{cr}(p_i)$, and $IL_{bn}(p_i)$. \\
$g_i$              & The $i^\mathrm{th}$ port group. \\
$w_{g_i}$& Check region of group-based congestion penalty. \\
$\lambda_c$        & The coefficient of group-based congestion penalty. \\
$s$        & Routing grid size. \\
\hline
\end{tabular}
}
\label{tab:notation}
\end{table}

%% file: figtex/fig_eocomparison.tex
\begin{figure}
    \centering
    \includegraphics[width=\columnwidth]{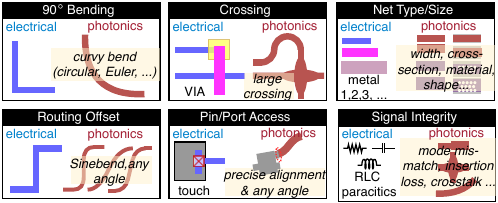}
    \caption{Compare properties/rules of EIC and PIC routing. 
    }
    \label{fig:EOComparison}
\end{figure}

%% file: doc/4_algo.tex
\section{\name: Automated PIC Detailed Routing}
\label{sec:Method}
In this section, we present the details of our proposed APR framework, built on a customized grid-based A$^\ast$ search algorithm.
It efficiently finds curvy waveguide paths and inserts crossings automatically to minimize maximum insertion loss while honoring design rules.
The overall flow of our proposed framework is shown in Fig.~\ref{fig:oview}. 
The core of our routing framework includes three main phases:
\ding{202}~\emph{Port Access Assignment}: This phase assigns ports, considering orientation and density, to ensure smooth routing and minimize congestion;
\ding{203}~
\emph{Iterative Curvy-Aware Waveguide Routing}: This phase connects all nets with curvy-aware A$^\ast$ search following group-based net ordering, with a local rip-up and reroute (LRR) check to optimize crossings and comply with the design rules in Section~\ref{sec:Preliminaries};
and \ding{204}~\emph{Route Refinement}: At the end of the routing stage, we refine the routing solution and generate a DRV-free GDS layout.
\input{figtex/fig_overview}

\subsection{Accessibility-Enhanced Port Assignment}
\label{sec:PortAssign}
The port access problem is one of the most challenging subroutines in PIC detailed routing. 
Unlike the VLSI routing problem, where metal pins are unidirectional, PICs use \emph{directional waveguide ports}, which have strict \textbf{access orientation and precise alignment requirements}.
Ports must be accessed with waveguides in a specific face-to-face orientation (180$^\circ$) and exact cross-section alignment, as shown in Fig.~\ref{fig:CheckOrientation}. 
Accessing the target port with a wrongly-oriented waveguide fails to find a legal connection, as there may not be enough space near the port to adjust direction using curvy bends.
When a waveguide passes near ports of other nets, accessing the \emph{blocked} port becomes even more difficult.
The primary reason for this port access challenge is the large area required to accommodate curvy waveguide bends.
To solve the above challenges, we propose the following port access assignment techniques that account for both port orientation and port density, enhancing overall port accessibility.
\input{figtex/fig_check_region}

\noindent\underline{\textbf{Port Propagation}}.~
In PICs, some ports are located within the device bounding box. Since devices are treated as obstacles, we propagate these internal ports to the boundary of the device bounding box according to their orientation, as shown in Fig.~\ref{fig:CheckOrientation}.

\noindent\underline{\textbf{Bending-Aware Port Access Region Reservation}}.~
To prevent other waveguides from blocking port regions, grids in front of each port, along the port orientation, are reserved for the corresponding net, ensuring they cannot be crossed by other nets, as shown in Fig.~\ref{fig:CheckOrientation}. 
The size of the reserved region is adaptive to the waveguide's bending radius, ensuring enough space for potential bends to maximize port access success while minimizing area.

\noindent\underline{\textbf{Congested Port Spreading}}.~
In some PIC devices, high-density ports may occupy the same routing grid, causing port access difficulty. 
To address this, we \emph{symmetrically} spread these access ports with a predefined extension length and spacing, as illustrated in Fig.~\ref{fig:PortSpreading}.
The newly arranged ports will connect to the original ports using sine bends, ensuring they \textbf{occupy distinct routing grids to reduce congestion}. 
The reserved port access region will be updated to reflect the new port locations.

\noindent\underline{\textbf{Channel Planning via Staggered Access Point Offsets}}.~
To enhance accessibility, we propose staggered access point regions for densely placed ports, as depicted in Fig.~\ref{fig:NetGroup}. 
For instance, in multimode interference (MMI) devices with numerous ports on the same side, high port density can lead to access ports being obstructed by nearby waveguides.
Parallel waveguides with narrow spacings prevent other nets from crossing over them, as inserting crossings requires sufficient space.
We \textbf{progressively extend the access region length} for inner ports with an offset larger than a waveguide crossing size. 
This approach not only leaves enough bending space for inner ports to navigate out of congested regions but also facilitates the \textbf{placement of consecutive crossings}, allowing other waveguides to pass through parallel waveguides.
This significantly decreases the chance of infeasible routing or excessive detours.

\subsection{Port-Group-based Net Order}
\name is a sequential router that processes nets one at a time.
The order of net routing impacts the final routing quality and feasibility.
We propose a port-group-based net ordering strategy that organizes ports on the same device based on their direction. 
Ports facing the same direction are clustered together into groups (e.g., $g_i$).
For example, as shown in Fig.~\ref{fig:NetGroup}, the 0° and 180° ports in a device are divided into two port groups $g_1$ and $g_2$. 
The \textbf{routing process is completed one group at a time}, ensuring that all nets within a group are routed before proceeding to the next group. 
The key insight behind this method is the observation that \emph{most congestion and routing conflicts arise between nets within the same group}. 
By employing a group-wise routing approach, \textbf{nets are routed with awareness of others in the same group, minimizing intra-group conflicts} and improving overall routing quality.
The routing order of a net $n_i$ is given by the following priority score $PR_{n_i}$ when a minimum-priority queue is used to manage all unrouted nets:
\begin{equation} 
\label{eq:netorder}
    PR_{n_i} = dist_{g}+ o_{n_i}=\text{min}_{n_i\in g}dist_{n_i}+o_{n_i}, n_i\in g
\end{equation}
where $dist_{n_i}$ is the Euclidean distance between two end ports of net $n_i$, and $dist_{g}$ denotes the smallest Euclidean distance among the nets within group $g$.
A smaller max net distance of a group will result in a higher routing order for that group.
The term $o_{n_i}$ refers to the local order of net $n_i$ within its group.
This ensures that nets are routed based on their relative position in the group, helping to reduce conflicts between nets, particularly in multiport devices.

\subsection{Non-Manhattan Waveguide Routing with Curvy-Aware A$^\ast$ Search}
In contrast to typical Manhattan VLSI routing, PIC designs typically employ non-Manhattan routing methods. 
Smooth curves decrease bending angles and waveguide lengths, thus reducing insertion loss.
In this section, we present our iterative waveguide routing algorithm, designed to generate smooth waveguides with both 45$^\circ$ and 90$^\circ$ turns, while supporting adaptive crossing insertion.

\subsubsection{Spacing-Ensured A$^\ast$ Routing Grid Size Setting}
The \name grid size $s$ is set to be larger than the waveguide width.
In typical PIC designs, waveguides are generally wider than ports. Setting $s$ larger than the waveguide width maximizes pathfinding efficiency while facilitating easy port access.

\subsubsection{Parametric Curvy-Aware Neighbor Candidate Generation}
\input{figtex/fig_neighbors}
To efficiently enable curvy-aware A$^\ast$ search, we propose parametric curvy-aware methods to generate neighbor candidates and perform comprehensive DRC check to select legal neighbors for exploration.

PIC uses curves instead of 90$^\circ$ 
 or 45$^\circ$ turns in VLSI and PCB routing.
We develop a customized curvy-aware neighbor generation scheme for the node based on parametric bending geometry. 
Each routing node is defined by its spatial location and orientation, which is crucial for accessing ports in the correct direction. 
We represent this as a \textbf{directional node} using ($x$, $y$, $orientation$).
As shown in Fig.~\ref{fig:NeighborDefinition}, we derive neighbor candidates based on the current node's orientation and a user-defined bending radius. 
Based on their orientation, current nodes are categorized into two states: the Manhattan State (MS) and the Non-Manhattan State (NMS).

The MS nodes align with the x/y-axis and have five neighbors: one adjacent neighbor at 0$^\circ$ and four non-adjacent neighbors at $\pm$45$^\circ$ and $\pm$90$^\circ$.
The NMS nodes are routed along the diagonal line with three neighbors.
The \textbf{location of neighbor candidates is adaptively derived} based on the bend radius ($\mathit{r}$) and grid size ($\mathit{s}$). 
Larger radii and smaller grids result in larger step sizes in the grid. 
For instance, for an MS node in 0$^\circ$, its adjacent neighbor is simply 1 grid away, and the steps of 90$^\circ$ and 45$^\circ$ neighbors are given by
\begin{equation} \label{eq:gcost}
\begin{aligned}
    & step_{90,x} = step_{90,y} = \ceil{r / s},  \\ 
    step_{45,x} = \ceil{(\sqrt{2}&-1) \cdot r/s};\quad step_{45,y} = \ceil{(1-\frac{\sqrt{2}}{2}) \cdot r/s}.
\end{aligned}
\end{equation}
In the neighbor generation process, we apply the ceiling function, $\ceil{\cdot}$, to ensure enough space for bending.
Unlike 45$^\circ$ diagonal neighbors in traditional 8-way A$^\ast$, where $step_{45,x}$ and $step_{45,y}$ are equal, our approach \textbf{intentionally sets $step_{45,x}$ to be larger than $step_{45,y}$, which prevents direct diagonal turns}. 
This is motivated by the fact that, at the corner grid, the 45$^\circ$ bend will indent inward toward the center, occupying the inner grid and disrupting the straight part before the corner. 
A larger $step_{45,x}$ ensures that it does not rely on the previously established search path and leaves enough space for the 45$^\circ$ bend.

\subsubsection{Geometry-Aware Neighbor Legality Check}
To ensure that only feasible neighbors are considered for exploration, a legality check is necessary before adding them to the priority queue. 
A neighbor is legal only when the real geometry of the corresponding waveguide does not violate any design rules.

\noindent\underline{\textbf{Hit No Obstacle: Geometry-Aware Spacing Check}}.~
If the neighbor does not hit an obstacle, we instantiate the real geometry of the connecting waveguide and perform a spacing check to ensure the route has no DRV.

\input{figtex/fig_crossing_insertion}

\noindent\underline{\textbf{Hit Routed Nets: Predictive Crossing Insertion}}.~
If a neighbor candidate hits a previously routed waveguide (marked as an obstacle), we need to check whether it is feasible to insert a waveguide crossing to pass through it.

As illustrated in Fig.~\ref{fig:CrossingInsertion}, several critical constraints must be considered for crossing insertion:
\ding{202}~\emph{Enough straight waveguide length}: 
Waveguide crossings occupy specific chip areas, requiring adequate spacing, and perpendicular orientations. 
Therefore, we check and ensure  sufficient straight length and correct port orientation by checking the orientation state at each routing grid.
\ding{203}~\emph{No conflict with blockages}:
We will check whether the bounding box of the CR overlaps with any obstacles to honor design rules.
\ding{204}~\emph{Port matching}: For successful connectivity, the waveguides must align precisely with the four ports of the crossing. 
This includes matching properties such as cross-section, width, etc.
By predictively checking all those legality conditions, we can adaptively incorporate crossing insertion during the routing process. 
This approach \textbf{reduces the need for long detours and avoids the complications associated with manually defined crossings} in the schematic.

\subsubsection{Insertion Loss-Aware A$^\ast$ Search Cost}
\input{figtex/fig_group_penalty}
\name uses a customized A$^\ast$ search cost to consider insertion loss and optimize the algorithm efficiency.
An A$^\ast$ search cost function $f(n)$ representing the cost of a path can be defined as $f(n) = g(n) + h(n)$,
where $g(n)$ is the cost from the source (s) to the current node $n$, and $h(n)$ is the estimated cost from the current node to the target $t$. 
The formulation of $g(n)$ is divided into two parts, the insertion loss of current node $g_{IL}(n)$ which follow the calculation of Eq.~\eqref{eq:iloss} and the \textbf{group-based congestion penalty (GCP)} $g_c(n, g_i)$:
\begin{equation} 
\small
\label{eq:gccost}
\begin{aligned}
    & g(n) = g_{IL}(n) + g_c(n, g_i), \\
    & g_c(n, g_i) = \lambda_c \cdot \#grids(w_{g_i}), \\ 
\end{aligned}
\end{equation}
where $\lambda_c$ is a penalty coefficient to \textbf{prevent the net from routing too close to the blockage or previously routed waveguides}, and $\#grids(w_{g_i})$ is the number of grids that occupied by others in the check region $w_{g_i}$ as shown in Fig.~\ref{fig:gcp}. 
$w_{g_i}$  is determined by the number of unrouted nets in its port group. 
As more nets are routed, $w_{g_i}$ decreases to avoid consuming extra space.
Empirically, $w_{g_i}$ aids the routing process by \textbf{reserving resources for each port group}, thereby preventing other nets from entering the port area.

We further customize the heuristic cost function $h(n)$ to better estimate 45$^\circ$ and 90$^\circ$ bends, as shown in Eq.~\eqref{eq:hcost}.
\begin{equation} 
\small
\label{eq:hcost}
\begin{aligned}
    & d_{min} = \text{min}(|n_x-t_x|, |n_y-t_y|), \\
    & d_{max} = \text{max}(|n_x-t_x|, |n_y-t_y|), \\
    & h(n) = d_{max} - d_{min} + \sqrt{2} \ast d_{min} + \alpha \cdot IL_{bd, 45}, \\
    & \alpha =
        \begin{cases} 
        1,  & \text{if } d_{min}>0,\ d_{max}>0 \\
        0,  & \text{others}
        \end{cases},
\end{aligned}
\end{equation}
where $d_{min}$ is the minimum difference between the current node $n$ and target node $t$ along either the x-axis or y-axis, and $d_{max}$ is the maximum one. 
The \textbf{insertion loss of 45$^\circ$ bend is added as a penalty} since a \textbf{non-zero $d_{min}$ or $d_{max}$ means there will be an orientation misalignment} in the end of the path and makes it hard to connect to the target port.

\subsubsection{Waveguide Instantiation}
\input{figtex/fig_waveguide_representation}
One of the largest differences of \name from prior global routing methods is \textbf{geometry awareness}. 
Once we obtain a path, we instantiate the curvy waveguide's real geometry with extrude function from GDSFactory~\cite{GDSFactory} and store it on the \textbf{overlapped oriented routing grid map} accordingly as shown in Fig.~\ref{fig:wg}.
Later, the A$^\ast$ search engine can thereby treat the existing routed waveguides as obstacles and consider waveguide spacing check and crossing insertion conveniently.

\subsubsection{Violated Net Removal.}
When accessing to the oriented target port is failed, \name apply a rip-up-and-reroute (RR) scheme.
We relax DRC checking and record nets that conflict with the established paths.
These nets are subsequently ripped up and rerouted in subsequent iterations. 
To avoid repeating the same routing results and to mitigate congestion, a history cost~\cite{liu2013nctu} is updated in the history map prior to net removal. 
Empirically, this history map-based negotiation process successfully resolves routing failures by balancing the demands of various nets.

\subsection{Crossing-Waveguide Optimization}
\input{figtex/fig_ripup_reroute}

We propose a local ripup-and-reroute (LRR) scheme to further balance the waveguide length and CRs.

\name adopts group-wise net routing order and incorporates group-based congestion penalties. 
However, this strategy may cause longer-path nets within a group to be routed first, thereby blocking nets in other port groups. 
Additionally, routing through congested areas and utilizing CRs often leads to divergent solutions, causing pathfinding to miss optimal routes. 
To address these issues, our LRR scheme performs both \emph{crossing-enabled} and \emph{crossing-disabled} routing attempts and selects the solution with lower IL.

The LRR evaluation is activated if a solution is found as shown in Fig.~\ref{fig:lrr}. 
If the current routing solution (RS) does not involve CRs, we will directly use it as the optimal path.
Otherwise, it will be ripped up in a later stage.
If CRs occur, possible reasons are (1) a blockage caused by another waveguide requiring a crossing, (2) a crossing chosen to bypass congestion, or (3) high propagation loss for non-crossing paths.
To verify these three possibilities, a crossing-disabled routing (NCS) is then activated. 
If NCS finds a path without using CRs, the insertion losses of CS and NCS are compared, and the lower-loss path is selected.
If NCS fails, it indicates the net is blocked. 
In this case, the blocking waveguide is assessed. 
If it has never been ripped up before, the blockage is likely caused by the group-based net order, and this blocking net will be ripped up, as it will not affect CR re-insertions in subsequent iteration. %
Our LRR strategy empirically optimizes the routing by \textbf{balancing long waveguides and CRs}.

\subsection{Routed Waveguide Refinement}
\input{figtex/fig_adjustment}
Since our grid-based routing method often results in the port center not aligning perfectly with the grid center, a slight offset can occur between the final path and the access port, as shown in Fig.~\ref{fig:adjustment}.
To resolve this, we adjust the initial and final segments of the waveguide path to align with the target device port, ensuring that the bend radius along the path remains unaffected. 
If this adjustment is not feasible, the waveguide will be connected to the port using a sine bend to maintain proper alignment.

%% file: figtex/fig_overview.tex
\begin{figure}[t]
    \centering
    \includegraphics[width=\columnwidth]{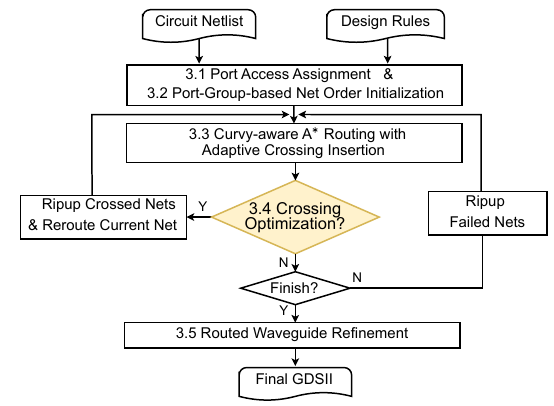}
    \vspace{-10pt}
    \caption{Algorithm flow of our \name framework.}
    \label{fig:oview}
    \vspace{-10pt}
\end{figure}

%% file: figtex/fig_check_region.tex
\begin{figure}
    \centering
    \vspace{-10pt}
    \subfloat[]{\includegraphics[width=0.5\columnwidth]{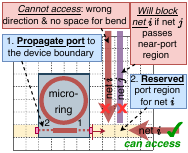}
    \label{fig:CheckOrientation}
    }
    \vspace{-10pt}
    \subfloat[]{\includegraphics[width=0.466\columnwidth]{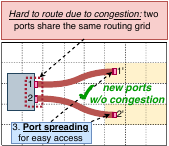}
    \label{fig:PortSpreading}
    }\\
    \subfloat[]{\includegraphics[width=\columnwidth]{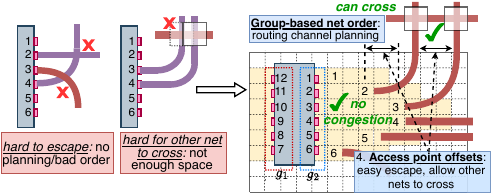}
    \label{fig:NetGroup}
    }
    \vspace{-10pt}
    \caption{(a) Port propagation and reserved port region help port access.
    (b) Port spreading removes congested ports in the same grid.
    (c) group-based net order with access point offset enables channel planning and allows potential crossing.}
    \label{fig:CheckRegion}
    \vspace{-5pt}
\end{figure}

%% file: figtex/fig_neighbors.tex
\begin{figure}
    \centering
    \includegraphics[width=0.93\columnwidth]{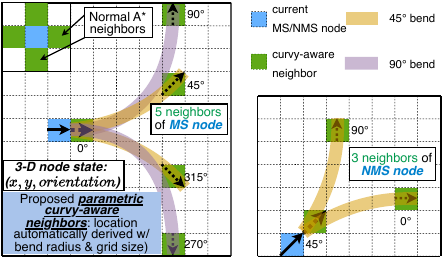}
    \vspace{-5pt}
    \caption{Parametric curvy-aware neighbors allow non-Manhattan curvy waveguide routing.
    Neighbors are automatically derived based on bending radius and grid size.
    }
    \label{fig:NeighborDefinition}
\end{figure}

%% file: figtex/fig_crossing_insertion.tex
\begin{figure}
    \centering
    \includegraphics[width=\columnwidth]{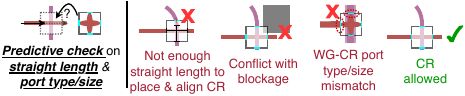}
    \vspace{-10pt}
    \caption{Proposed adaptive waveguide crossing insertion.}
    \label{fig:CrossingInsertion}
    \vspace{-5pt}
\end{figure}

%% file: figtex/fig_group_penalty.tex
\begin{figure}[t]
    \centering
    \includegraphics[width=8.5cm]{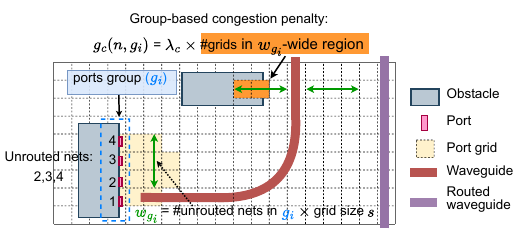}
    \vspace{-10pt}
    \caption{Group-based congestion penalty in Eq.~\eqref{eq:gccost}.}
    \label{fig:gcp}
    \vspace{-5pt}
\end{figure}

%% file: figtex/fig_waveguide_representation.tex
\begin{figure}[t]
    \centering
     \vspace{-5pt}
    \includegraphics[width=0.82\columnwidth]{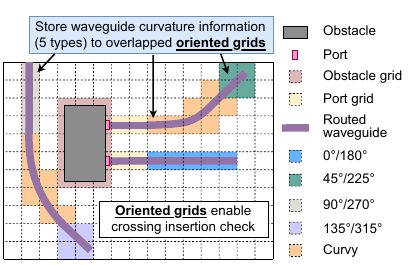}
    \vspace{-5pt}
    \caption{Represent routed waveguides in oriented grid map.}
    \label{fig:wg}
    \vspace{-5pt}
\end{figure}

%% file: figtex/fig_ripup_reroute.tex
\begin{figure}[t]
    \centering
    \includegraphics[width=\columnwidth]{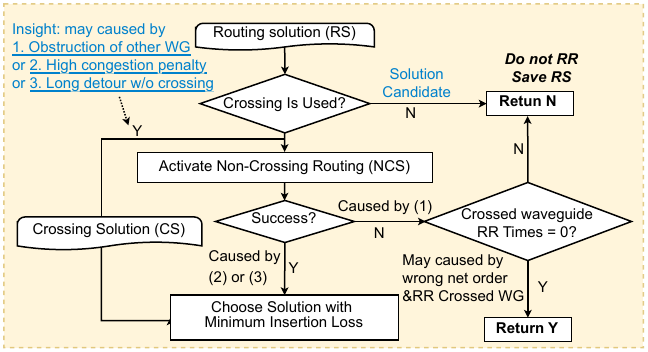}
    \vspace{-15pt}
    \caption{LRR check after finding a routing solution.}
    \label{fig:lrr}
    \vspace{-5pt}
\end{figure}

%% file: figtex/fig_adjustment.tex
\begin{figure}[t]
    \centering
    \includegraphics[width=0.8\columnwidth]{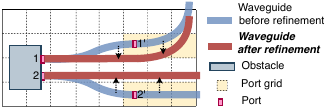}
    \vspace{-5pt}
    \caption{Waveguide refinement to remove bending.}
    \label{fig:adjustment}
    \vspace{-5pt}
\end{figure}

%% file: doc/5_result.tex
\section{Experimental Results}
\label{sec:ExperimentalResults}

\subsection{Experimental Setting}

The proposed photonic detailed routing framework is implemented in Python based on GDSFactory~\cite{GDSFactory}
libraries. 
All experiments are conducted on a personal workstation with an Intel i5-125600KF 3.7GHz CPU with 32GB memory.

\noindent\underline{\textbf{Benchmarks}}.~To assess the scalability of our proposed framework, we conduct experiments on different types of benchmarks: Photonic Tensor Cores (PTC) and Wavelength-routed Optical Network-on-Chip (WRONoC). 

PTCs and WRONoCs have very different characteristics.
Table~\ref{tab:bench_stat} shows the benchmark statistics. 
PTC circuits have a more structured topology but have limited routing resources and high port density.
For PTCs, we evaluate \name on Clements-style Mach-Zehnder interferometer (MZI) array~\cite{NP_Optica2018_Clements} and auto-searched PTC \texttt{ADEPT}~\cite{NP_DAC2022_Gu} with different scales. 
The bend radius is set by 5 $\mu m$ for single-mode Si waveguides (width=500 nm). 
WRONoC circuits, on the other hand, occupy a large die area and have unstructured interconnection topology. 
For WRONoCs, we conduct experiments on optical router benchmarks~\cite{truppel2019psion} with all the optical switches centered in the layout. 
Based on the positions of the memory controllers, we have four cases for this benchmark. 
The bend radius is set by $60 \mu m$ for its huge routing resource. 
Placement solutions of all benchmark circuits are designed manually by an experienced designer and verified with simulation using GDSFactory and KLayout.
To evaluate the critical path IL, we summarize the device IL used in Table~\ref{tab:parameters}.
\input{tables/tab_parameters}

\input{tables/tab_benchmark}

\noindent\underline{\textbf{Baselines}}.~We compare our \name with a prior method PROTON~\cite{proton}. 
Note that the original PROTON mainly focuses on path planning and crossing optimization with adaptive crossing penalty, which cannot generate real waveguide geometry.
For a fair comparison, we adapt PROTON by adding reserved port regions and a global ripup and reroute scheme to make it applicable to PIC detailed routing problems.
Two variants of the adapted PROTON are: (1) the original implementation with global RR scheme (\textbf{Base-1}) and (2) additional 45-degree bend neighbors with more rip-up and re-route iterations to address accessing problem (\textbf{Base-2}).

\subsection{PIC Routing Quality Evaluation}
\label{sec:MainResults}

\input{tables/tab_result}

We compare \name with PROTON~\cite{proton} in terms of critical path insertion loss $IL_{max}$, the critical path length, the number of crossings on the critical path, design rule violations (DRV), and wall-clock runtime.
Table~\ref{tab:stat} shows that our \name can generate \textbf{DRV-free layouts} on all benchmarks with an average of \textbf{14\% lower critical path IL and 6.25$\times$ speedup}.

\input{figtex/fig_mmi}

\noindent\underline{\textbf{Analysis of PTC Results}}.~The PTC benchmarks, featuring limited routing resources and high port density, provide a strong validation for a router's ability to place bends and crossings while successfully accessing the target ports.
(1) The Clements-style MZI array features a highly structured mesh topology with no inherent topological crossings, but suffers from non-ideal placement issues such as misalignments, flipped devices, and limited routing space. 
Due to the stringent routing spaces to access ports, baselines introduce extra waveguide CRs and lead to DRVs.
In contrast, our \name can find crossing-optimal (\#CR=0), DRV-free paths in much shorter runtime.
(2) \texttt{ADEPT} PTC is even more challenging due to the high port density in multi-port MMI devices and numerous topological crossings. 
As the size of the PTC increases, baselines exhibit a sharp rise in DRV and runtime.
\name shows superior \textbf{scalability}, consistently producing \textbf{DRV-free low-IL} layouts for large circuits with \textbf{2-22.5$\times$ faster runtime}. 
Figure~\ref{fig:MMI16} visualizes the DRV-free ADEPT\_16$\times$16 layout generated by \name with real curvy waveguide geometry and instantiated crossings.

\input{figtex/fig_noc}
\noindent\underline{\textbf{Analysis of WRONoC Results}}.~
WRONoC features a large chip area and unstructured interconnection topology, which makes it challenging for a router to explore the large search space.
It is important to note the counter-intuitive trade-off between \#CR and WL. 
In NoC benchmarks, where the die size is large, fewer CRs do not necessarily result in lower IL. 
Reducing CRs may cause considerably longer detours, increasing propagation loss and ultimately leading to a higher overall $IL_{max}$.
Aside from the case Router\_oneside, our \name exhibits the minimum $IL_{max}$ across the remaining cases with crossing-optimal (\#CR=0), DRV-free layout.

\subsection{Discussion}
\label{sec:Ablation}
\input{tables/tab_ablation}

\noindent\underline{\textbf{Non-Manhattan 45-Degree Routing}}. Compared to \textbf{Base-1}, \textbf{Base-2} achieves an average of 19\% shorter critical path WL by introducing the 45-degree bend (diagonal neighbors), which validates the effectiveness of a non-Manhattan routing style in PICs.

\noindent\underline{\textbf{Crossing-Disabled Routing (NCS)}}.~
As shown in Table~\ref{tab:stat} (\textbf{Base-2} vs. \name), our proposed additional crossing-disabled routing trial (NCS) introduces an extra runtime penalty, but it reduces the overall runtime and leads to higher solution quality as it \emph{mitigates the port access issue and leads to much fewer total RR iterations}.

\noindent\underline{\textbf{Port-Group-based Congestion Penalty (GCP)}}.~We evaluate the benefits of our proposed group-based congestion penalty in optimizing crossings using CRs with different IL: high crossing IL with $\alpha_c=1$ and low crossing cost with $\alpha_c=0.3$ as shown in Table~\ref{tab:ablation}. 
When crossings have high IL, our method effectively avoids crossings, minimizing the maximum insertion loss ($IL_{max}$). For low-IL crossings, it opts for paths with shorter WL with more CRs to optimize $IL_{max}$.
Without the group-based penalty, however, the algorithm turns out to increase CR usage as $\alpha_c$ rises, as it struggles to find a low-\#CR path due to congestion from other nets. 
By applying our group-based penalty, \textbf{net conflicts are largely reduced}, enabling more efficient routing decisions with fewer crossings and lower IL.
As shown in Fig.~\ref{fig:NOC}, our $\alpha_c$ factor serves as a flexible control knob, enabling users to adjust crossing insertion according to their preferences and specific PIC performance requirements, such as phase balancing and reduced crosstalk.

%% file: tables/tab_parameters.tex
\begin{table}
\centering
\caption{Device IL parameters used in $IL_{max}$ evaluation.
}
\vspace{-5pt}
\resizebox{8.5cm}{!}{
\begin{tabular}{|c|c|c|c|c|c|}
\hline
Propagation $\alpha_w$     & Bending $\alpha_b$ & CR $\alpha_c$ & Y-branch & MZI & MMI \\\hline
1.5 dB/cm  ~\cite{chan2010architectural}      & 0.005 dB ~\cite{chan2010architectural} & 0.52 dB~\cite{proton} & 0.3 dB~\cite{ybrand} & 1.2 dB~\cite{mzi} & 0.1 dB~\cite{mmi} \\
\hline
\end{tabular}
}
\label{tab:parameters}
\vspace{-5pt}
\end{table}

%% file: tables/tab_benchmark.tex
\begin{table}
\centering
\caption{Benchmark PIC information. 
}
\vspace{-5pt}
\resizebox{8.5cm}{!}{
\begin{tabular}{|c|c|c|c|c|c|}
\hline
Benchmark & \#Devices & \#Nets & Die Size & Waveguide Width & Grid Size\\ \hline
Clements\_8$\times$8~\cite{NP_Optica2018_Clements}           &52          &79           & 4800$\times$1600 $\mu m^2$     & 0.5 $\mu m$ & 0.2 $\mu m$\\ 
Clements\_16$\times$16 ~\cite{NP_Optica2018_Clements}        &168         &287           & 8000$\times$3200 $\mu m^2$    & 0.5 $\mu m$ & 0.2 $\mu m$\\ 
\texttt{ADEPT}\_8$\times$8~\cite{NP_DAC2022_Gu}               &82          &111            & 4400$\times$1600 $\mu m^2$  & 0.5 $\mu m$ & 0.2 $\mu m$\\ 
\texttt{ADEPT}\_16$\times$16~\cite{NP_DAC2022_Gu}              &162         &223           & 6900$\times$3200 $\mu m^2$   & 0.5 $\mu m$ & 0.2 $\mu m$\\ 
\texttt{ADEPT}\_32$\times$32~\cite{NP_DAC2022_Gu}              &318         &446            & 13000$\times$6400 $\mu m^2$  & 0.5 $\mu m$ & 0.2 $\mu m$\\ 
Routers~\cite{zheng2021topro}                      &9         &16           & 10000$\times$10000 $\mu$m$^2$    & 2 $\mu m$ & 50 $\mu m$\\ \hline
\end{tabular}
}
\label{tab:bench_stat}
\vspace{-5pt}
\end{table}

%% file: tables/tab_result.tex
\begin{table*}[t]
\centering
\caption{Comparisons of the maximum insertion loss value $IL_\mathit{max}$ (dB), the path length with $IL_\mathit{max}$ (WL ($\mu m$)), the number of crossings passed by the signal with $IL_\mathit{max}$, total design rule violations (DRV), and runtime (s). 
$\downarrow$: lower is better.}
\vspace{-5pt}
\resizebox{17.8cm}{!}{
\begin{tabular}{cccccccccccccccc}
\hline
& \multicolumn{5}{c}{Base-1 (Adaptive crossing penalty)~\cite{proton} }                                                                       & \multicolumn{5}{c}{Base-2 (w/ Diagonal neighbors)~\cite{proton}}                                                                                   & \multicolumn{5}{c}{\textbf{\name}}                                                                                                                 \\
\cmidrule(lr){2-6} \cmidrule(lr){7-11} \cmidrule(lr){12-16}
\multirow{-2}{*}{Benchmark} & \#CR & WL (mm) & \cellcolor[HTML]{D9D9D9}$IL_{max}\downarrow$ (dB)    & \cellcolor[HTML]{D9D9D9}DRV $\downarrow$ & \cellcolor[HTML]{D9D9D9}Time $\downarrow$ (s)       & \#CR & WL (mm)       & \cellcolor[HTML]{D9D9D9}$IL_{max}\downarrow$ (dB)    & \cellcolor[HTML]{D9D9D9}DRV $\downarrow$ & \cellcolor[HTML]{D9D9D9}Time $\downarrow$ (s)    & \#CR & WL (mm) & \cellcolor[HTML]{D9D9D9}$IL_{max}\downarrow$ (dB)     & \cellcolor[HTML]{D9D9D9}DRV $\downarrow$        & \cellcolor[HTML]{D9D9D9}Time $\downarrow$ (s)         \\
\hline
Clements\_8x8               & 0    & 3.39    & \cellcolor[HTML]{D9D9D9}16.99      & \cellcolor[HTML]{D9D9D9}0   & \cellcolor[HTML]{D9D9D9}113      & 1    & 3.01          & \cellcolor[HTML]{D9D9D9}16.82      & \cellcolor[HTML]{D9D9D9}0   & \cellcolor[HTML]{D9D9D9}258         & 0    & 2.94    & \cellcolor[HTML]{D9D9D9}16.38         & \cellcolor[HTML]{D9D9D9}0          & \cellcolor[HTML]{D9D9D9}32            \\
Clements\_16x16             & 5    & 5.06    & \cellcolor[HTML]{D9D9D9}29.32      & \cellcolor[HTML]{D9D9D9}12  & \cellcolor[HTML]{D9D9D9}580      & 2    & 4.34          & \cellcolor[HTML]{D9D9D9}27.52      & \cellcolor[HTML]{D9D9D9}4   & \cellcolor[HTML]{D9D9D9}274         & 0    & 4.38    & \cellcolor[HTML]{D9D9D9}26.74         & \cellcolor[HTML]{D9D9D9}0          & \cellcolor[HTML]{D9D9D9}164           \\
ADEPT\_8x8                  & 16   & 4.7     & \cellcolor[HTML]{D9D9D9}17.12      & \cellcolor[HTML]{D9D9D9}26  & \cellcolor[HTML]{D9D9D9}179      & 18   & 4.16          & \cellcolor[HTML]{D9D9D9}17.46      & \cellcolor[HTML]{D9D9D9}17  & \cellcolor[HTML]{D9D9D9}249         & 18   & 4.1     & \cellcolor[HTML]{D9D9D9}18            & \cellcolor[HTML]{D9D9D9}0          & \cellcolor[HTML]{D9D9D9}98            \\
ADEPT\_16x16                & 28   & 7.84    & \cellcolor[HTML]{D9D9D9}24.07      & \cellcolor[HTML]{D9D9D9}98  & \cellcolor[HTML]{D9D9D9}1306     & 17   & 7.66          & \cellcolor[HTML]{D9D9D9}18.36      & \cellcolor[HTML]{D9D9D9}26  & \cellcolor[HTML]{D9D9D9}2627        & 16   & 7.38    & \cellcolor[HTML]{D9D9D9}17.8          & \cellcolor[HTML]{D9D9D9}0          & \cellcolor[HTML]{D9D9D9}243           \\
ADEPT\_32x32                & 66   & 16.13   & \cellcolor[HTML]{D9D9D9}44.57      & \cellcolor[HTML]{D9D9D9}355 & \cellcolor[HTML]{D9D9D9}9981     & 52   & 13.97         & \cellcolor[HTML]{D9D9D9}37.19      & \cellcolor[HTML]{D9D9D9}181 & \cellcolor[HTML]{D9D9D9}27140       & 50   & 15.04   & \cellcolor[HTML]{D9D9D9}36.34         & \cellcolor[HTML]{D9D9D9}0          & \cellcolor[HTML]{D9D9D9}1204          \\
router\_north               & 6    & 32.98   & \cellcolor[HTML]{D9D9D9}11.09      & \cellcolor[HTML]{D9D9D9}0   & \cellcolor[HTML]{D9D9D9}36       & 6    & 21.63         & \cellcolor[HTML]{D9D9D9}9.37       & \cellcolor[HTML]{D9D9D9}0   & \cellcolor[HTML]{D9D9D9}66          & 0    & 31.11   & \cellcolor[HTML]{D9D9D9}7.78          & \cellcolor[HTML]{D9D9D9}0          & \cellcolor[HTML]{D9D9D9}81            \\
router\_oneside             & 0    & 18.71   & \cellcolor[HTML]{D9D9D9}5.89       & \cellcolor[HTML]{D9D9D9}0   & \cellcolor[HTML]{D9D9D9}5        & 4    & 20.96         & \cellcolor[HTML]{D9D9D9}8.26       & \cellcolor[HTML]{D9D9D9}0   & \cellcolor[HTML]{D9D9D9}33          & 0    & 21.55   & \cellcolor[HTML]{D9D9D9}6.31          & \cellcolor[HTML]{D9D9D9}0          & \cellcolor[HTML]{D9D9D9}37            \\
router\_corner              & 8    & 20.81   & \cellcolor[HTML]{D9D9D9}10.23      & \cellcolor[HTML]{D9D9D9}1   & \cellcolor[HTML]{D9D9D9}54       & 9    & 16.7          & \cellcolor[HTML]{D9D9D9}10.1       & \cellcolor[HTML]{D9D9D9}0   & \cellcolor[HTML]{D9D9D9}75          & 0    & 35.29   & \cellcolor[HTML]{D9D9D9}8.4           & \cellcolor[HTML]{D9D9D9}0          & \cellcolor[HTML]{D9D9D9}59            \\
router\_pairwise            & 7    & 28.49   & \cellcolor[HTML]{D9D9D9}10.94      & \cellcolor[HTML]{D9D9D9}1   & \cellcolor[HTML]{D9D9D9}48       & 8    & 19.52         & \cellcolor[HTML]{D9D9D9}10.05      & \cellcolor[HTML]{D9D9D9}0   & \cellcolor[HTML]{D9D9D9}66          & 0    & 33.52   & \cellcolor[HTML]{D9D9D9}8.14          & \cellcolor[HTML]{D9D9D9}0          & \cellcolor[HTML]{D9D9D9}65            \\
\hline
Geo-mean                    & -    & 15.34   & \cellcolor[HTML]{D9D9D9}18.91      & \cellcolor[HTML]{D9D9D9}-   & \cellcolor[HTML]{D9D9D9}1367     & -    & 12.44         & \cellcolor[HTML]{D9D9D9}17.24      & \cellcolor[HTML]{D9D9D9}-   & \cellcolor[HTML]{D9D9D9}3421        & -    & 17.26   & \cellcolor[HTML]{D9D9D9}16.21         & \cellcolor[HTML]{D9D9D9}-          & \cellcolor[HTML]{D9D9D9}220           \\
Ratio                       & -    & 1       & \cellcolor[HTML]{D9D9D9}1          & \cellcolor[HTML]{D9D9D9}-   & \cellcolor[HTML]{D9D9D9}1        & -    & \textbf{0.81} & \cellcolor[HTML]{D9D9D9}0.91       & \cellcolor[HTML]{D9D9D9}-   & \cellcolor[HTML]{D9D9D9}2.5         & -    & 1.12    & \cellcolor[HTML]{D9D9D9}\textbf{0.86} & \cellcolor[HTML]{D9D9D9}\textbf{-} & \cellcolor[HTML]{D9D9D9}\textbf{0.16} \\
\hline
\end{tabular}
}
\label{tab:stat}
\end{table*}

%% file: figtex/fig_mmi.tex
\begin{figure}
    \centering
    \includegraphics[width=0.93\columnwidth]{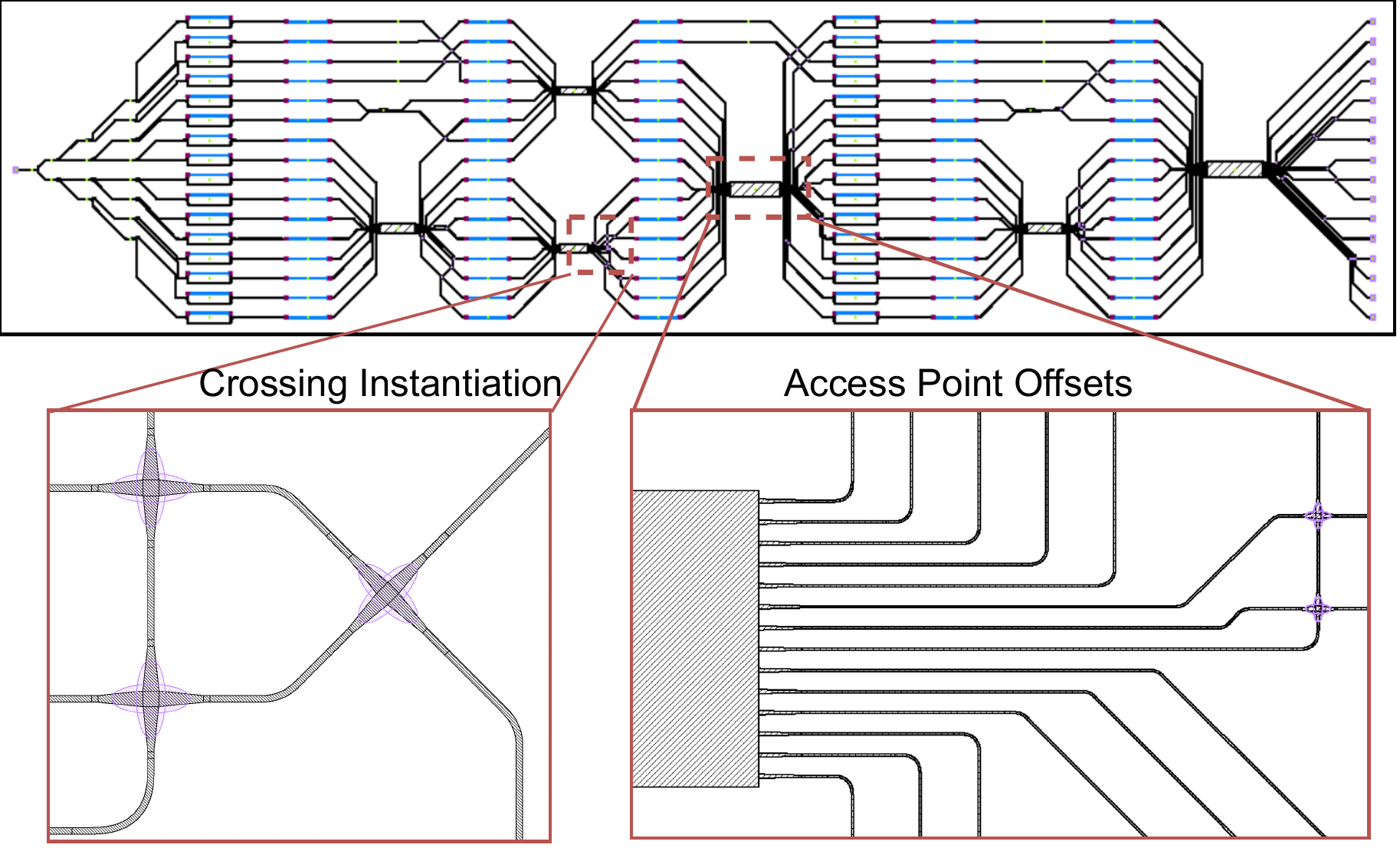}
    \vspace{-5pt}
    \caption{Layout of \texttt{ADEPT}\_16$\times$16~\cite{NP_DAC2022_Gu} routed by our \name.}
    \vspace{-5pt}
    \label{fig:MMI16}
\end{figure}

%% file: figtex/fig_noc.tex
\begin{figure}
    \centering
    \includegraphics[width=\columnwidth]{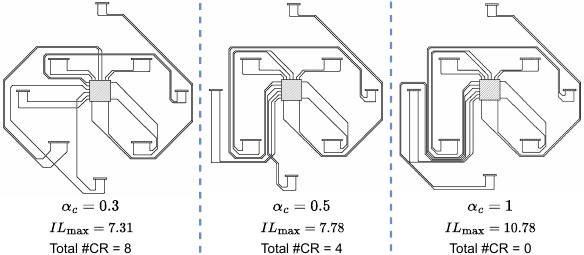}
    \vspace{-10pt}
    \caption{Layout of router\_north of different crossing loss.
    }
    \label{fig:NOC}
    \vspace{-5pt}
\end{figure}

%% file: tables/tab_ablation.tex
\begin{table}
\centering
\caption{Ablation study of GCP with different crossing cost.}
\resizebox{8cm}{!}{
\begin{tabular}{ccccc}
\hline
\multirow{2}{*}{Metrics} & \multicolumn{2}{c}{High Crossing Cost $\alpha_c=1$}    & \multicolumn{2}{c}{Low Crossing Cost $\alpha_c=0.3$}   \\  
\cmidrule(lr){2-3} \cmidrule(lr){4-5}
                           & w/o GCP        & \textbf{\name}           & w/o GCP        & \textbf{\name} \\       \hline
\#CR                     & 6                      & 0                   & 5                      & 5                    \\
WL (mm)                       & 20.72                  & 31.11               & 25.11                  & 26.04                \\
\cellcolor[HTML]{D9D9D9}$IL_{max}\downarrow$                    & \cellcolor[HTML]{D9D9D9}15.21                  & \cellcolor[HTML]{D9D9D9}10.78               & \cellcolor[HTML]{D9D9D9}7.18                   & \cellcolor[HTML]{D9D9D9}7.31                 \\
\cellcolor[HTML]{D9D9D9}DRV                      & \cellcolor[HTML]{D9D9D9}0                      & \cellcolor[HTML]{D9D9D9}0                   & \cellcolor[HTML]{D9D9D9}1                      & \cellcolor[HTML]{D9D9D9}0                    \\
\cellcolor[HTML]{D9D9D9}Time (s)                     & \cellcolor[HTML]{D9D9D9}129                  & \cellcolor[HTML]{D9D9D9}73                & \cellcolor[HTML]{D9D9D9}261                  & \cellcolor[HTML]{D9D9D9}197  \\ 
\hline
\end{tabular}
}
\label{tab:ablation}
\end{table}

%% file: doc/6_conclu.tex
\section{Conclusion}
\label{sec:Conclusion}
We introduce \name, an open-source automated detailed routing tool specifically designed for photonic integrated circuits (PICs). 
\name features a non-Manhattan curvy-aware A$^\ast$ search engine with accessibility-enhanced port assignment, adaptive crossing insertion, congestion-aware group-based net ordering and objective, and crossing-waveguide optimization scheme to handle unique PIC routing constraints while optimizing critical path insertion loss.
On large-scale PIC benchmarks, \name demonstrates its capability to generate DRV-free layouts with 14\% lower insertion loss and a 6.25$\times$ speedup compared to prior approaches, which highlight \name's potential to significantly advance EPDA for complex photonic systems, paving the way for more efficient, scalable PIC designs.